\documentclass[twocolumn]{revtex4}
\usepackage{epsfig}

\newcommand{\be}{\begin{equation}}
\newcommand{\ee}{\end{equation}}
\newcommand{\bea}{\begin{eqnarray}}
\newcommand{\eea}{\end{eqnarray}}
\newcommand{\di}{\partial}

\newcommand{\nd}{\noindent}
\newcommand{\gm}{\gamma_{\mu}}
\newcommand{\gn}{\gamma_{\nu}}
\newcommand{\ta}{\gamma_{35}}
\newcommand{\psr}{\bar{\Psi}}

\newcommand{\chib}{\bar{\chi}}

\begin{document}

\title{Chiral symmetry breaking in the ${\rm QED}_{3}$ in
 presence of irrelevant interactions:
\\a renormalization group study}
\author{Kamran Kaveh and Igor F. Herbut}

\affiliation{Department of Physics, Simon Fraser University,
Burnaby, British Columbia, \\ Canada V5A 1S6\\}

\begin{abstract}
Motivated by recent theoretical approaches
to high temperature superconductivity, we study dynamical
mass generation in three dimensional quantum electrodynamics
(${\rm QED}_{3}$) in presence of irrelevant four-fermion quartic
terms. The problem is reformulated in terms of the renormalization
group flows of certain four-fermion couplings and charge, and then 
studied in the limit of large number of fermion flavors $N$. We
find that the critical number of fermions $N_c$ below which the
mass becomes dynamically generated depends continuously on a weak
chiral-symmetry-breaking interaction. One-loop calculation in our
gauge-invariant approach yields $N_{c0} = 6$ in pure ${\rm
QED}_3$. We also find that chiral-symmetry-preserving mass cannot
become dynamically generated in pure ${\rm QED}_{3}$.
\end{abstract}
\maketitle \vspace{10pt}

\section{Introduction}
It has been proposed recently that the low-energy theory of
gapless quasiparticles in a two-dimensional d-wave superconductor
(dSC) with strong phase fluctuations can be represented by the
two-flavor massless quantum electrodynamics in three dimensions
(${\rm QED}_{3}$) \cite{FT1}, \cite{IgorPRL}. The coupling
constant (or the `charge') in such an effective theory is the
vortex condensate, i.~e. the order parameter dual to the  usual 
superconducting order parameter. It is well known that 
${\rm QED}_{3}$ is inherently unstable towards the dynamical mass
generation \cite{Pisarski} \cite{Appelquist}, which in the context
of d-wave superconductivity implies the transition into one of
several possible insulating ground states. Each of the
insulators corresponds to a broken generator of the ${\rm U}(4)$
{\it chiral} symmetry of ${\rm QED}_{3}$, which emerges at low
energies in the  standard dSC \cite{IgorPRL}, \cite{FT4}. Most
important among the insulating ground states is the
spin-density-wave, which turns out to be favored by the repulsive
interactions \cite{IgorPRL}, \cite{Babakh}. This approach provides
then a viable unified description of the known low-temperature phases
of underdoped high-temperature superconductors.

Dynamical mass generation, however, occurs only if the number of
Dirac fermions $N$ in ${\rm QED}_{3}$ does not exceed the
critical number $N_{c0}$. If the value of $N_{c0}$ turns out to be
less than the number of Dirac fermions, which for a single-layer
dSC is $N = 2$, then quantum disordering of the phase of the dSC
will yield a spin liquid, instead the spin-density-wave insulator.
It is thus of importance to establish whether ${\rm QED}_{3}$
with $N=2$ lies below or above the critical value for 
spontaneous chiral symmetry breaking in the theory.

The estimates of $N_{c0}$ at the moment strongly disagree, however.
Early studies of Schwinger-Dyson equations in large-N
approximation gave $N_{c0}=32/\pi^2\approx 3.24$ \cite{Appelquist}.
Vertex corrections \cite{Maris}, or
the next-to-leading-order terms in the $1/N$ expansion
\cite{Nash} did not change $N_{c0}$
significantly, and if anything, only increased its value. On the other hand,
Appelquist {\it et. al.} have argued that $N_{c0}<3/2$
\cite{Appelquist3}. Adding to the controversy, recent lattice calculations
have found no decisive signal for chiral symmetry breaking for $N=2$,
but did detect a significant fermion mass for $N=1$ \cite{lattice}.
It has been argued, however, that although greatly increased
compared to early studies, the sizes of
the systems considered in the lattice calculations may still not be
close enough to the thermodynamic limit \cite{Gusynin}. In fact, due to the
essential singularity at $N=N_c$ the value of the mass at $N=2$, if
finite, should be rather small, and the results of numerical simulations are
not necessarily in conflict with the values obtained from the
Schwinger-Dyson equations \cite{lattice}, \cite{mavromatos}.

In the context of high-temperature superconductivity, however, an
additional issue arises. Chirally symmetric, Lorentz invariant
${\rm QED}_{3}$ emerges only asymptotically at low energies, when
all the irrelevant perturbations may be ignored. For example,
large anisotropy between the two characteristic velocities of
the dSC, although marginally irrelevant \cite{vafek}, \cite{dlee},
\cite{thomas}, 
reduces the ${\rm U}(4)$ symmetry of the two-flavor theory to the
${\rm U}(2)\otimes{\rm U}(2)$ over a wide crossover region
\cite{IgorPRL}. The (irrelevant) repulsive
interaction between electrons
breaks each ${\rm U(2)}$ factor per flavour further down to
${\rm U}(1)\otimes{\rm U}(1)$. It is presently unclear how, and if
at all, the presence of these irrelevant perturbations affects the
value of $N_c$ in the more complete theory.
This is the issue we wish to address in the present paper.

We apply the momentum-shell renormalization group (RG) to 
${\rm QED}_{3}$ theory with $N$ fermion flavours, and with
four-fermion interactions which break the ${\rm U(2)}$ symmetry per
flavour. The gauge-invariant $\beta$-functions for the charge and  the 
four-fermion couplings are computed to the leading order in $1/N$.
The value of $N_{c}$ may be obtained from the RG flow simply by
inverting the dependence of the critical coupling(s) $g$ on $N$.
In case of symmetry breaking interaction we show that $N_c$ obtained this
way is necessarily a
monotonic function of the interaction coupling, i. e. that an 
infinitesimal interaction, although irrelevant, alters the value
of $N_c$. In particular, this suggests that even if $N_{c0} < 2$
in pure ${\rm QED}_{3}$, the low-energy theory of underdoped
cuprates with repulsive interactions included \cite{IgorPRL} is
likely to lie below the (shifted) critical point for dynamical
mass generation. The flow of the chirally-symmetric interactions,
on the other hand,  suggests that the chirally symmetric mass cannot
get spontaneously generated in pure ${\rm QED}_{3}$.

 Our method relies on identification of the RG runaway flow of the
chiral-symmetry-breaking interaction coupling constant with the
dynamical mass generation. This conjecture is supported by the
exact solution in the limit $N=\infty$ and of zero charge.
 The idea is rather general, however, and similar to the standard way of
determining a spontaneously broken symmetry in statistical
physics: first allow a weak explicit symmetry breaking
perturbation, take the thermodynamic limit, and only then take the
perturbation to zero. Thermodynamic limit would in the RG language
correspond to letting the momentum cutoff go to zero.

The article is organized as follows: In Sec.~\ref{quartic} we
introduce the symmetry-breaking and the symmetry-preserving
 four-fermion interactions. In Sec.~\ref{RGlanguage} we formulate the
 problem of dynamical mass generation in the  RG language. In
Secs.~\ref{CSBSec} and~\ref{CSPSec} the RG flows are derived in
the full theory with all the important  quartic interactions taken
into account. Concluding remarks are given in Sec. VI, and some 
technical details are presented in the Appendix.


\section{QED$_3$ and quartic interactions}
\label{quartic}

We begin by reviewing briefly
the spin sector of the low-energy theory of the phase-disordered d-wave
superconductor \cite{IgorPRL}, described by the action
$S=\int d^3 x L$, with the Lagrangian
\bea
L = L_{{\rm QED}_{3}} + L_{{\rm int}} +
L_{{\rm hd}},\nonumber\\L_{{\rm QED}_{3}} =
\bar{\Psi}_{i}\gm(\di_{\mu} + ia_{\mu})\Psi_{i} +
\frac{1}{2e^{2}} (\nabla \times {\bf a})^{2}.
\label{action0}
\eea
$\Psi_{i}$, $i=1,2$ represent
the electrically neutral  spin-1/2 fermions (spinons),
$\psr = \Psi^{\dagger}\gamma_{0}$,
$\gm$'s are the usual Dirac gamma matrices ($\mu
= 0,1,2$), and we define
$\gamma_{5} = \gamma_{0}\gamma_{1}\gamma_{2}\gamma_{3}$,
and $\gamma_{35} = i\gamma_{3}\gamma_{5}$  for later use \cite{IgorPRL}.
The charge $e^2 \sim |\langle \Phi \rangle|^2$, where $\Phi$ is
the vortex loop condensate
\cite{IgorPRL}, \cite{leespin}. The complementary
charge sector of the theory may be shown to be describing an insulator
\cite{igorunpub}.

The short-range
repulsive interaction may be written in terms of Dirac fermions as
\bea
L_{{\rm int}} &=&
U(i\psr_{i}\gamma_{5}\gamma_{1}\Psi_{i})^{2}. \label{Hubbard}
\eea
Higher derivatives in the kinetic energy, similarly, take the form
\be L_{{\rm hd}} \sim \psr_{i}
\gamma_{5}[\gamma_{1}f(\partial^{2}) -
\gamma_{2}g(\partial^{2})]\Psi_{i}. \label{NL}\ee
where the functions $f(z)$ and $g(z)$ come from the expansion of the
quasiparticle dispersion near the nodes \cite{IgorPRL}.

The velocity anisotropy neglected in Eq.(1) in principle
reduces the full ${\rm U}(4)$ symmetry of ${\rm QED}_{3}$ to
${\rm U}(2) \otimes{\rm U}(2)$. Each
${\rm U}(2)= {\rm U}(1)\otimes {\rm SU}_c (2)$
factor is generated by the algebra
$\{{\bf 1}, \gamma_{3} , \gamma_{5}, \ta\}$
where ${\rm SU}_{c}(2)$ is the chiral symmetry (CS) subgroup generated
by the last three generators. Inclusion of $L_{{\rm int}}$ and $L_{{\rm hd}}$
reduces the ${\rm SU}_c (2)$ symmetry further down to the ${\rm
U}_{c}(1)$ generated by $\gamma_{5}$, which is simply the
generator of translations in the nodal direction in this language.
Since the mass that turns out to be dynamically generated in $S$ is
$m\sim \langle\bar{\Psi}\Psi\rangle$, which preserves
$\gamma_{35}$ \cite{Babakh}, it will prove more
convenient to consider interactions
that directly preserve that particular generator.

Let us first consider the case of single fermion species, and
then generalize to $N>1$. To construct the quartic interaction
that breaks the ${\rm SU}_c(2)$ symmetry down to ${\rm U}_c (1)$
we notice that the three-component objects, \bea
{\bf A} = (\psr\Psi, ~~ \psr i\gamma_{3}\Psi, ~~ \psr i\gamma_{5}\Psi),\nonumber\\
{\bf B}_{\mu} = (\psr\gm\ta\Psi, ~~ \psr i\gm\gamma_{3}\Psi, ~~ \psr
i\gm\gamma_{5}\Psi),
\nonumber\\
\eea
are the only triplets under the chiral group.
Upon breaking the symmetry to ${\rm U}_{c}(1)$, we look at the
projection of {\bf A} and ${\bf B}_{\mu}$ along the  direction
corresponding to the remaining generator of the ${\rm SU}_{c}(2)$.
In this case, these are $\psr\Psi$ and $\psr\gm\ta\Psi$ which remain
invariant under the action of $\ta$. Thus, the required quartic
chiral-symmetry-breaking (CSB) interaction will have the
form
\be L_{{\rm CSB}} = \frac{g}{N}(\bar{\Psi}\Psi)^{2} +
\frac{g^{\prime}}{N}(\bar{\Psi}\gm\ta\Psi)^{2}. \label{CSBint} \ee

On the other hand, the two ${\rm SU}_c (2)$ singlets \be C_\mu =
\psr\gm\Psi, ~~~~~~~ C_{35} = \psr\ta\Psi, \ee \nd may be used to
construct the chiral-symmetry-preserving (CSP)  quartic
interactions, as \be L_{{\rm CSP}} =
\frac{\lambda}{N}(\bar{\Psi}\ta\Psi)^{2} +
\frac{\lambda^{\prime}}{N}(\bar{\Psi}\gm\Psi)^{2}.\\
\label{CSPint} \ee

For a general $N$ we will therefore define the following
${\rm U} (N)\otimes {\rm U}(N)$ symmetric theory
\bea L &=& L_{{\rm QED_{3}}} + L_{{\rm CSB}} + L_{{\rm CSP}},\nonumber\\
 &=& \{\bar{\Psi}_{i}\gm (\di_{\mu} + ia_{\mu})\Psi_{i}
+ \frac{1}{2 e^2} (\nabla \times {\bf a})^2
+ \frac{g}{N}(\bar{\Psi}_{i}\Psi_{i})^2
\nonumber\\&+& \frac{g^{\prime}}{N}(\bar{\Psi}_{i}\gm\ta\Psi_{i})^2
+\frac{\lambda}{N}(\bar{\Psi}_{i}\ta\Psi_{i})^2+
\frac{\lambda^{\prime}}{N}(\bar{\Psi}_{i}\gm\Psi_{i})^2\},\nonumber\\
&i& = 1\cdots N. \label{action} \eea

In principle, one could imagine other interaction terms
satisfying the required symmetry. However, it can be shown that these would
have to be a linear combination of the already introduced quartic terms. For
example, the (`Nambu-Jona-Lasinio') interaction
$g_{1}|{\bf A}|^2 + g_{2}|{\bf B}_{\mu}|^2$ can be written as a linear
combination of $C_\mu ^2$ and $C_{35} ^2$. This follows from 
Fierz identities which imply that there
are only two linearly independent quartic terms invariant under the
${\rm U}(2N)$. For ${\rm U}(N)\otimes{\rm U}(N)$ theory, the number of
independent couplings doubles to four, which
are precisely the introduced $g,~g^{\prime},~\lambda~{\rm and}~\lambda^{\prime}$.
For a more detailed discussion we refer the reader to the Appendix.

In the next section we focus on a single four-fermion interaction and
try to understand the spontaneous chiral symmetry breaking within the
 renormalization group approach.


\section{Dynamical mass generation in the RG language}
\label{RGlanguage}

An exactly solvable case of the theory in Eq.
(\ref{action}) is in the limit of infinite number of fermion flavours
and of zero charge ($e=0$). Let us first consider a single CSB
interaction term, $(g/N)(\psr\Psi)^2$, and set $g' =
\lambda=\lambda'=0$
 (i. e. the Gross-Neveu model). For $N\rightarrow \infty$,
such interaction gives rise to a dynamically generated mass, $m \sim
\langle \psr\Psi \rangle$, determined by the gap equation
\be - \frac{1}{g} =
8\int \frac{d^{3}k}{(2\pi)^{3}}\frac{1}{k^2 + m^2},
\label{gap} \ee
\nd which after the integration gives
\be 1 =
\frac{4 g \Lambda}{\pi^{2}}(\frac{m}{\Lambda}\tan^{-1}\frac{\Lambda}{m}-1 ),
\ee
with $\Lambda \gg m $ being the assumed ultraviolet (UV) cutoff.
Demanding $m$ to be invariant under the change of cutoff
 $\Lambda\rightarrow \Lambda/b$,
the $\beta$-function at $N=\infty$ is readily obtained to be exactly
\be
\beta_g = \frac{dg}{d\ln b} = -g - g^{2},
\label{gapRG}
\ee
where $g$ has been rescaled as $ 4g\Lambda/ \pi^{2} \rightarrow g$.
We see that a weak coupling $g$ is irrelevant, but that the flow for
$g<g_* = -1$, which represents the infrared (IR) unstable fixed point,
 is towards negative infinity. Since the same values of $g$ yield a finite
 mass from the gap equation, it is natural to
 identify the runaway flow of $g$ with the dynamical mass generation.
Note that the same Eq. (\ref{gapRG}) can alternatively be
obtained in the standard Wilson's momentum-shell one-loop RG.

The second solvable limit of the theory is
pure ${\rm QED}_{3}$ without any four-fermion interactions,
again in the limit $N\rightarrow \infty$. The flow of the charge is then
\be \beta_e = \frac{de^{2}}{d\ln b} = e^{2} - Ne^{4}, \ee \nd where the
dimensionless charge is defined as $(4/3)(e^2 /(2\pi^2\Lambda))
\rightarrow e^2$. While the theory is free in the UV region, there is
a non-trivial IR stable fixed point at $e_* ^2 = 1/N$. (Notice that
the quartic interactions, even when present,
can not appear in $\beta_e$ to the leading
order in large $N$ as a consequence of Ward-Takahashi identity.)

\begin{figure}[h]
\begin{center}
\epsfig{figure=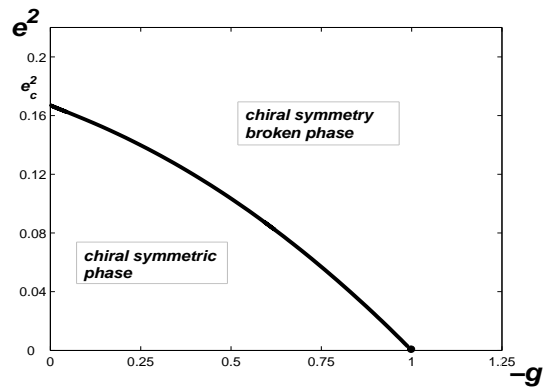, height=150pt, width=220pt,angle=0}
\end{center}
\caption{The phase-diagram in the interaction-charge plane, for the chiral symmetry
breaking interaction. The value of the charge is 
$e^2 = 1/N$. $N_c $ is a continuous function of the symmetry breaking
interaction, as a consequence of the existence of the fixed point at
$g=0$ at any charge.}
\label{eg}
\end{figure}

Next, we want to consider the interplay of the charge $e$ and the quartic
coupling $g$, and in particular to examine the influence of a weak
charge on the value of $g_*$. One expects the effect of the gauge
field on $\beta_{g}$ to be
\be \frac{dg}{d\ln b} = -g -
g^{2}\nonumber + ({\rm const.})~e^2g, \label{RGintuitive}\ee \nd
to the leading order in $e^2$ and $1/N$. In particular, $\beta_g
(g=0)=0$ even when $e\neq 0$, since otherwise it would be possible
to {\it generate} the CSB interaction in the chirally symmetric
theory. So $g=0$ is {\it always} a fixed point. Since at the fixed
point $e_* ^2 = 1/N$, decreasing $N$ is the same as increasing the
charge in Eq. (13). Since the factor in front of the last term is
expected to be positive (as it indeed turns out to be the case),
decreasing $N$ will reduce the absolute value of the non-trivial 
critical coupling $g_*$, until it
eventually merges with the trivial fixed point, permanently
located at $g=0$. There will therefore exist a critical charge $e_{c}^2
= 1/N_{c0}$, above which an {\it infinitesimal} symmetry breaking
interaction suffices to cause the runaway flow of $g$.  We
identify this point with the {\it spontaneous} chiral symmetry
breaking in pure ${\rm QED}_{3}$. The flow diagram with this
structure has been depicted in Fig \ref{eg}.

When $N<\infty$ the terms with  an
explicit $N$-dependence in $\beta_{g}$, such as $g^3/N$, should also be
included. These terms may be understood as contributing
the $1/N$ corrections to $N_{c0}$ in the following way.
One can expand the critical charge
(corresponding to the double-root of $\beta_g$ at $g= 0$)
in powers of $1/N$ as
\be
e^2_{c} = a_{0} + \frac{a_{1}}{N_{c0}} + \frac{a_{2}}{N_{c0} ^2} + \cdots,
\label{selfconsistent}.
\ee
Since $e^2_{c} =
1/N_{c0}+ O(1/N_{c0} ^2 )$ from $\beta_e$, this effectively generates then
the $1/N$-expansion for $N_{c0}$.

One may analogously consider the CSP interaction
$(\lambda/N)(\psr\ta\Psi)^2$, which when alone leads to the
dynamically generated mass $m\sim\langle  \psr\ta\Psi \rangle$ for
$\lambda < -1$, in the $N\rightarrow\infty$ limit. In presence of
the charge, however, there is a crucial difference between the
$\beta_\lambda$ and $\beta_g$. Since the CSP interaction term has
the same full chiral symmetry as the pure ${\rm QED}_{3}$, finite
charge may, and in fact does, {\it generate} the coupling $\lambda$. This
manifests itself as the $e^4$ contribution in $\beta_\lambda$, which
will now take the form \be
\frac{d\lambda}{d\ln b} = -\lambda - \lambda^2 + ({\rm const.})~
e^2\lambda + ({\rm const.})~e^4. \ee With the last term, however,
 $\lambda=0$ {\it is not} a fixed point any longer.
Furthermore, the sign of the $e^4$-term turns out to be {\it
positive}, so that the critical coupling actually {\it increases}
with charge. We interpret the latter feature
as that the spontaneous dynamical
generation of the chiral symmetry preserving mass in pure ${\rm
QED}_{3}$ is not possible. This would be in agreement with 
conclusions of the earlier studies  \cite{AppelquistParity},
\cite{Vafa}.


\section{RG for CSB interactions and the value of $N_{c0}$}
\label{CSBSec}

In general, the $\beta$-functions for
all four quartic interactions will be coupled and the flow is non-trivial.
To the leading order in $1/N$, however, the
calculation simplifies considerably. In the following two theorems
we show that the $\beta$-functions for the CSB and CSP
interactions are completely decoupled in this limit.

{\bf Theorem I}: To the leading order in $1/N$ and
for $e = 0$, different $\beta$-functions decouple.

Proof: Only particle-hole diagrams, as in Fig. (\ref{feynman1})
contribute to leading order in large $N$.
Such diagrams are proportional to:

\begin{figure}[h]
\begin{center}
\epsfig{figure=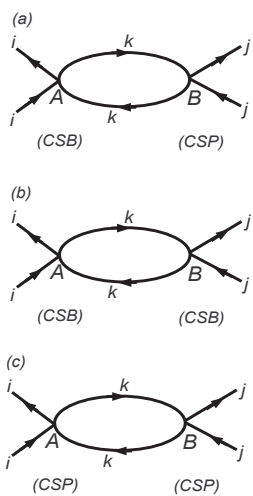, height=200pt,
width=100pt,angle=0}
\end{center}
\caption{Particle-hole diagrams to the leading order in $1/N$.}
\label{feynman1}
\end{figure}

\bea
\propto g_{A}g{_B}\cdot
\int d^{3}q~{\rm Tr}(\Gamma_{A}G(q)\Gamma_{B}G(q)) 
\propto {\rm Tr}(\Gamma_{A}\Gamma_{B}). \label{trace}
\eea
Here, $\Gamma_{A}$ and $\Gamma_{B}$'s are the matrices in the kernel of
the quadratic form accompanying either $g_{A}$ or $g_{B}$,
\be
\Gamma_{A}, \Gamma_{B} \in \{ {\bf 1}, \gm,
\ta\gm, \ta \}
\ee
It is easy to see that these diagrams are zero unless $g_{A} =
g_{B}$. For diagrams that mix CSB and CSP interactions in
Fig. (\ref{feynman1}a), Eq. (\ref{trace}) contains the trace of an odd
number of $\gamma$-matrices and thus yields zero. For
the CSB-CSB or CSP-CSP diagrams in Figs. (\ref{feynman1}b) and
(\ref{feynman1}c), the identities ${\rm
Tr}(\gm\gn)=4\delta_{\mu\nu}$ and ${\rm Tr}(\gm\ta) = 0$,
imply that all the mixing terms are zero unless $\mu = \nu$,
i.~e. $g_{A} = g_{B}$.

So, to the leading order, the coupling between
different quartic interactions in the $\beta$-functions
can only be mediated through charge.
One may easily see, however, that the symmetry requires that the
$\beta$-functions for the CSB and CSP couplings
still remain decoupled. We will state it in
the form of the following theorem:

{\bf Theorem II:} There are no $\sim \lambda e^2$ or $
\sim \lambda^{\prime}e^{2}$ terms in $\beta_{g}$ or
$\beta_{g^{\prime}}$, nor $\sim ge^2$ and $\sim g^{\prime}e^2$ terms in
$\beta_{\lambda}$ and $\beta_{\lambda^{\prime}}$, to the leading order
in $1/N$.

Proof: $\lambda e^2$ and $\lambda^{\prime} e^2$ terms obey the full chiral
symmetry, and thus cannot generate a CSB interaction. To prove the
equivalent statement for the CSB couplings, we notice that to the leading
order in $1/N$ the $g e^2$ and $g' e^2$ terms differ from the
$\lambda e^2$ and $\lambda^{\prime} e^2$ terms by a single
$\gamma_{35}$ matrix, and thus necessarily
break the chiral symmetry. They therefore cannot generate a CSP coupling.
\begin{figure}[h]
\begin{center}
\epsfig{figure=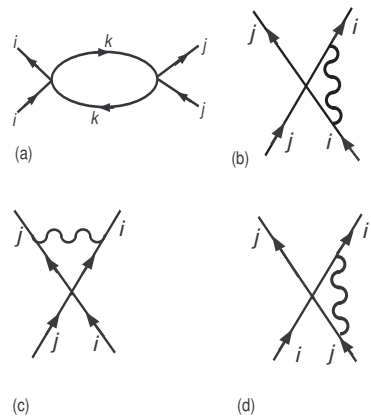, height=160pt,
width=140pt,angle=0}
\end{center}
\caption{Diagrams contributing to the renormalized couplings to
the leading order in $1/N$.} \label{feynman2}
\end{figure}
\begin{figure}[h]
\begin{center}
\epsfig{figure=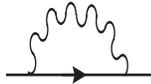, height=40pt,
width=60pt,angle=0}
\end{center}
\caption{Diagram contributing to the wavefunction renormalization.}
\label{wavefunctionRG}
\end{figure}

The above theorems allow us to significantly  reduce the number of
relevant Feynman diagrams. The straightforward calculation of the
diagrams in Figs. 3 and 4 leads to the following $\beta$-functions for the
symmetry breaking interactions:
\bea \frac{de^{2}}{d\ln b} &=& e^{2} - Ne^{4}\nonumber\\
\frac{dg}{d\ln b} &=& -g - g^2 + 4e^{2}g +
18e^{2}g^{\prime}\nonumber\\
 \frac{dg^{\prime}}{d\ln b} &=& -g^{\prime} + g^{\prime 2}+
\frac{2}{3}e^{2}g,
\label{CSB}
\eea
with conveniently rescaled parameters \nd
\be
4 g\Lambda /\pi^{2}\rightarrow g,
4g^{\prime}\Lambda /(3\pi^{2}) \rightarrow g^{\prime},
2e^{2}/(3\pi^{2}\Lambda) \rightarrow e^{2}\nonumber\\.
\ee
Note that the coupling $g'$ becomes generated by $g$ and $e$
even if absent initially, so in
principle it must be included into the analysis.
A notable feature of the above $\beta$-functions is also their independence
on the gauge-fixing parameter $\xi$. This derives from the exact
cancellation between the
gauge-dependent part of the diagrams in Fig. 3
and the wavefunction renormalization factor $Z$ (Fig. 4):
\be Z = 1 + (\xi -\frac{2}{3})e^2 \ln b.
\ee

 The flow diagram on the $g-g'$ plane for $N=\infty$ $(e^2=0)$
is given in Fig. 5.  For $N < \infty$, the fixed point value of
the charge becomes $e^2 = 1/N$, and the locations of all the fixed
points, except the trivial one at the origin, shift in the
directions as indicated. The point at which the RG trajectory that starts at 
the purely repulsive fixed point (initially at $(-1, 1)$) and terminates at
the `Gross-Neveu' fixed point (initially at $(-1,0)$) intersects the
$g$-axis determines the location of the phase boundary in the $g-e^2$  
($g'=0$) plane. At small charge we obtain such a phase boundary at 
\begin{equation}
g= -1 + 4 e^2 +O(e^4), 
\end{equation}
whereas at low $g$ 
\begin{equation}
g= -\frac{144}{13} (\frac{1}{6} - e^2) + O( (\frac{1}{6} - e^2)^2 ).
\end{equation}
Numerical solution at a general coupling is
given at Fig. 1.  The critical point in pure
${\rm QED}_{3}$, $N_{c0}$, is determined by the value of $N$ for
which  `Gross-Neveu' fixed point reaches the origin.
For $N>N_{c0}$, the flow beginning at an infinitesimal negative
$g$ and $g'=0$ then runs away to infinity. To the leading order in
$1/N$, this criterion yields $N_{c0} = 6$. At $N=N_{c0}$ the
other two non-trivial fixed points are still at finite values. The role of
$g'$ is therefore only to modify the phase boundary in the
$g-e^2$ plane and the value of $N_{c0}$ quantitatively, but not
qualitatively.  Neglecting the flow of $g'$ entirely would lead, for example,
to $N_{c0}=4$. This would correspond to the value at which the dimension of
the coupling $g$ at the charged fixed point changes sign. 

\begin{figure}[h]
\begin{center}
\epsfig{figure=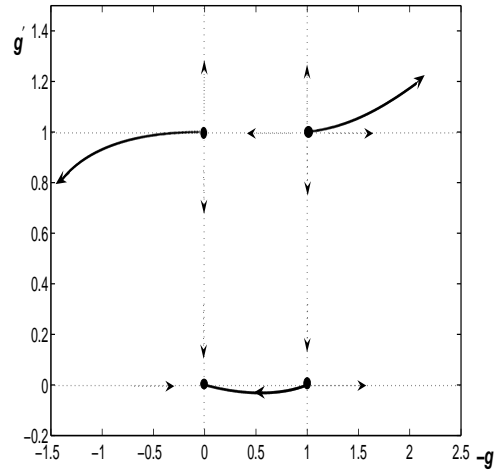, height=200pt, width=200pt,angle=0}
\end{center}
\caption{The RG flow (dashed lines) in the plane of
the two symmetry-breaking interactions $g$ and $g^{\prime}$ for
$N=\infty$. Full lines
mark the evolution of the four fixed points with decrease of $N$.}
 \label{matlab1}
\end{figure}

\section{RG for CSP interactions}
\label{CSPSec}

We now turn to the analysis of the theory in Eq. 8 with $g=g'=0$,
i. e. when the quartic terms respect the full chiral symmetry.
Although somewhat artificial from the point of view of the
effective theory  for underdoped cuprates, this exercise
underlines the  important role of symmetry in the phase diagram.
The diagrams are still the same as in the CSB case, with the
addition of the two diagrams in Fig. 6. These new terms {\it
generate} the coupling $\lambda$, and thus change the evolution of
the flow diagram with $N$ in an important way, as mentioned in the
introduction and depicted in Fig. 7.
We obtain the following $\beta$-functions for the
couplings $\lambda, \lambda^{\prime}$ and $e^{2}$: \bea
\frac{de^{2}}{d\ln b} &=& e^{2} - Ne^{4}\nonumber\\
\frac{d\lambda}{d\ln b} &=& -\lambda - \lambda^{2} +
4e^{2}\lambda + 18e^{2}\lambda^{\prime} + 9 N e^4\nonumber\\
\frac{d\lambda^{\prime}}{d\ln b} &=& -\lambda^{\prime} +
\lambda^{\prime 2} + \frac{2}{3}e^{2}\lambda. \label{CSP} \eea
Notice that the flow equations for the CSB and CSP cases are
identical, apart from the positive $e^4$ term. This term, however, prevents
the fixed point that was
located at $(-1,0)$ for $N=\infty$ to ever merge with the
Gaussian fixed point, and consequently, no spontaneous generation
of the chiral-symmetry-preserving mass should be
 allowed in pure ${\rm QED}_{3}$.

\begin{figure}[h]
\begin{center}
\epsfig{figure=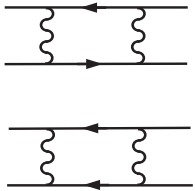, height=80pt, width=80pt,angle=0}
\end{center}
\caption{The diagrams that give $e^4$ term in $\beta_{\lambda}$.}
 \label{e4}
\end{figure}

\begin{figure}[h]
\begin{center}
\epsfig{figure=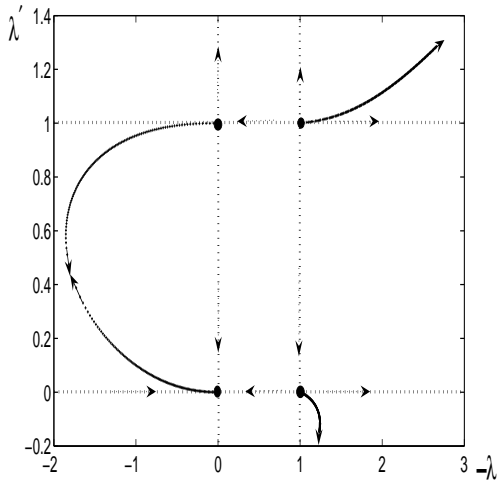, height=200pt, width=200pt,angle=0}
\end{center}
\caption{The evolution of the chiral symmetry preserving fixed points
in the $\lambda-\lambda^{\prime}$ plane with the increase of charge.}
\label{matlab2}
\end{figure}


\section{Conclusion}

 In conclusion, by reformulating the problem of dynamical mass generation
in ${\rm QED}_{3}$ with four-fermion interactions in terms of
the renormalization group flows, we found that the critical number
of fermions $N_c$ is a continuous function of the
chiral-symmetry-breaking interaction. By taking the limit of
vanishing interactions we estimated that $N_{c0}=6$ in pure
${\rm QED}_{3}$. Our analysis of the chiral-symmetry-preserving
interactions suggests that the chiral-symmetry-preserving mass cannot
become dynamically generated in pure ${\rm QED}_{3}$.

The result that the $N_c$ may depend on  an infinitesimal symmetry
breaking interaction should be contrasted with the previous
studies of Schwinger-Dyson equations in ${\rm QED}_{3}$ with
symmetry preserving interactions (the `gauged Nambu-Jona-Lasinio
model'). There, the $N_c$ was found to depend on the quartic
interaction only if the latter is larger than a certain value
\cite{miransky}. In the RG language this would correspond to the
merger of the two fixed points, like the Gaussian and the
`Gross-Neveu' fixed points
 in our case, at a {\it finite} value of the
coupling. In fact, we find that occurring in the Eqs. 21 for CSP
interactions: the `Gaussian' fixed point (initially at $(0,0)$) and
the `Thirring' fixed point (initially at $(0,1)$) for $N=4.83$ meet
at $(1.78,0.43)$. For $N>4.83$ both couplings become complex, and
the flow that begins at the line $\lambda=0$ is always towards
infinite $\lambda'$. It is tempting to identify this runaway flow
with the phase with broken chiral symmetry and the dynamically
generated mass, as proposed in \cite{Terao2}. We refrain from
doings so, however, since the runaway flow for $\lambda ' > 1$ at
$e=\lambda=0$ (the `Thirring model'), actually {\it does not}
correspond to the broken symmetry phase, as one can easily check
by directly solving the gap equation in this case at $N=\infty$.
The transition in the Thirring model occurs only at the order of
$1/N$ \cite{hong}, and so we suspect that the above runaway flow of $\lambda'$
may be an artifact of the $N=\infty$ limit. This issue is left for
a future study.

  Finally, although our scheme provides a systematic way of computing
$N_{c0}$, for example, it becomes rapidly complicated.
 To the next order in $1/N$ CSP and CSB coupling constants mix in the
 $\beta$-functions. Since the couplings $\lambda$ and $\lambda'$ get
 generated by the charge, and then mix into $\beta_g$,
 one necessarily has to track the flow of all four couplings.

\acknowledgements{}

This work was supported by NSERC of Canada. K. K. thanks B. Seradjeh for
helpful discussions.


\section{Appendix: Fierz identities and
generality of the interaction Lagrangian}

In this appendix we will construct the linear relationship between
the quartic terms invariant under a unitary ${\rm U}(N)$ group,
known as Fierz identities. These are the direct consequences of
the completeness relation for the generators of the symmetry
group.

 Defining ${\rm Tr}(A\cdot B)$ as the inner product between matrices $A$ and $B$,
we write down the completeness relation for the basis constructed
out of generators of a ${\rm U}(N)$ group, $\{\lambda^{\alpha},
{\bf 1}\}$, as

\be
\frac{1}{N}\delta_{ab}\delta_{cd} +
\frac{1}{2}\sum_{\alpha}^{N^2-1}\lambda^{\alpha}_{ab}
\lambda^{\alpha}_{cd} = \delta_{ad}\delta_{cb}.
\label{flavour}
\ee

As the special case of ${\rm U}(2)$, using Pauli matrices, Eq.
(\ref{flavour}) simplifies to \be \delta_{ab}\delta_{cd} +
\sum_{\alpha}\sigma^{\alpha}_{ab} \sigma^{\alpha}_{cd} =
2\delta_{ad}\delta_{cb}. \label{spinor} \ee Using the above
relations one can derive the requisite linear relationship between
different quartic terms. First, it is convenient to represent the
$4N$-component spinor $\Psi$ in terms of $2N$-component ones as
\be
\Psi=\left(\begin{array}{c}\chi^{i}\\\phi^{i}\end{array}\right).
\ee It is then possible to apply to above completeness relations
to the quartic terms of the form \be
\sum_{\alpha}(\chib^{i}_{a}\lambda^{\alpha}_{ij}\chi^{j}_{a})
(\bar{\chi}^{k}_{b}\lambda^{\alpha}_{kl}\chi^{l}_{b}), \ee where
$\chi^{i}$ stands for both $\chi^{i}$ and $\phi^{i}$. (The spinor
index is indicated by subscripts.) Applying Eq. (\ref{spinor}) for
spinor degrees of freedom and Eq. (\ref{flavour}) for flavour
degrees of freedom, one ends up with the following identity \be
(1+\frac{1}{N})(\bar{\chi}\chi)^{2} +
\sum_{\mu}(\bar{\chi}\sigma_{\mu}\chi)^{2} +
\sum_{\alpha}(\bar{\chi}\lambda^{\alpha}\chi)^{2} = 0,
\label{fierz1} \ee where we have suppressed both the spinor and
flavour (large-$N$) indices for convenience, and replaced $N$ with
$2N$, since ${\rm QED}_{3}$ is ${\rm U}(2N)$-symmetric. Similarly,
beginning with the quartic term \be
\sum_{\alpha,\mu}(\bar{\chi}^{i}_{a}\lambda^{\alpha}_{ij}\sigma^{\mu}_{ab}\chi^{j}_{b})(\bar{\chi}^{k}_{c}
\lambda^{\alpha}_{kl}\sigma^{\mu}_{cd}\chi^{l}_{d}), \ee it is
easy to derive the other identity \cite{gomes}
\bea \sum_{\alpha,
\mu}(\bar{\chi}\lambda^{\alpha}\sigma_{\mu}\chi)^{2} +
\sum_{\alpha}(\bar{\chi}\lambda^{\alpha}\chi)^{2} +
\frac{1}{N}\sum_{\mu}(\bar{\chi}\sigma_{\mu}\chi)^{2}\nonumber\\
+ (4 + \frac{1}{N})(\bar{\chi}\chi)^{2} = 0. \label{fierz2} \eea

The above identities applied to a ${\rm U}(2N)$-symmetric theory
with the $S_{{\rm int}}$ of the form \be
\tilde{g}_{1}(\chib\chi)^2 +
\tilde{g}_{2}(\chib\lambda^{\alpha}\chi)^2 +
\tilde{g}_{3}(\chib\sigma_{\mu}\chi)^2 +
\tilde{g}_{4}(\chib\sigma_{\mu}\lambda^{\alpha}\chi)^2,
\label{chi} \ee leave only two of the terms as independent.
Noticing that the Eq. (\ref{chi}) is equivalent to the interaction
term written in $4N$-component representation: $g_{1}|{\bf A}|^2 +
g_{2}|{\bf B}_{\mu}|^2 + g_{3}|C_{\mu}|^2 + g_{4}|C_{35}|^2$, we
can see that our choice of $C_{\mu}$ and $C_{35}$ as the most
general CSP quartic terms is justified.

 CSB case is not very different. One can consider the interaction
of the form in Eq. (\ref{chi}) for each ${\rm U}(N)$ sector
separately (i.e. $\chi$ and $\phi$). Repeating the same argument
would reduce the number of independent interaction couplings in each
sector to two, so that the
 overall number of independent couplings will be four, as assumed
in  Eq. (\ref{action}).

\end{document}